\documentclass[11pt, a4paper]{article}
\usepackage{jcappub}
\usepackage[utf8]{inputenc}
\usepackage[T1]{fontenc}
\usepackage{graphicx}

\usepackage[numbers]{natbib}
\usepackage{aas_macros}

\def\tsc#1{\csdef{#1}{\textsc{\lowercase{#1}}\xspace}}
\tsc{WGM}
\tsc{QE}

\newcommand{\rhob}{\rho_b}
\newcommand{\rhoDM}{\rho_{\rm DM}}
\newcommand{\Pb}{P_b}
\newcommand{\PDM}{P_{\rm DM}}

\begin{document}
\let\WriteBookmarks\relax
\def\floatpagepagefraction{1}
\def\textpagefraction{.001}

\title{Self-Interacting Dark Matter in Brown Dwarfs}
\author[a]{A. P\'erez-Garrido}
\affiliation[a]{Departamento de F\'\i sica Aplicada, Universidad Polit\'ecnica de Cartagena, Cartagena, Spain}
\emailAdd{antonio.perez@upct.es}

\abstract{Brown dwarfs, being transitional objects between giant planets and low-mass stars, possess dense, cool interiors that
provide optimal conditions to explore non-standard physics.
Capture and accumulation of dark-matter particles can alter the thermal,  structural and dynamic  of these substellar objects.
We aim to apply a self-consistent two-fluid framework to model the internal structure of self-gravitating brown dwarfs and to
quantify how the presence of a dark-matter component modifies their mass--radius relations and dynamical properties.
The brown dwarf is modeled as a composite system of a baryonic fluid, described by a polytropic equation of state,
and an independent dark-matter fluid. Both components are coupled through their shared gravitational
potential in hydrostatic equilibrium. We solve numerically the coupled Lane-Emden equations for a range of dark-matter mass
fractions.
We find that dark matter accumulating in the core reshapes the baryonic density profile, modifying both the radius and
the second-order Love number.
Radius and dynamical anomalies in brown dwarfs can serve as diagnostic tools to constrain dark-matter
properties. Future high-precision astrometric missions could identify these structural signatures, establishing
brown dwarfs as possible detectors of dark matter in the Galaxy.}


\arxivnumber{2605.XXXXX}
\maketitle

\section{Introduction}
Brown dwarfs were observationally confirmed three decades ago \cite{rebolo1995,nakajima1995} 
as a distinct class of substellar objects bridging the gap between giant planets and 
hydrogen-burning stars. Their internal structure is well described by partially or fully 
degenerate equations of state \cite{auddy2016}, which naturally give rise to a nearly flat mass--radius 
relation over a wide mass range. 
As a result, brown dwarfs are highly sensitive to gravitational perturbations, making them targets to 
probe for non-baryonic components.
Even a subdominant non-baryonic contribution can 
induce modifications to their density profiles, radii, and tidal responses, 
with  small changes in their overall masses.

The nature of dark matter (DM) remains the most compelling evidence for physics beyond the Standard Model (SM). While the Cold Dark
Matter (CDM) paradigm effectively explains large-scale structure formation, the microphysical identity of the DM particle persists
as an open question. For decades, the theoretical landscape was dominated by Weakly Interacting Massive Particles (WIMPs) in the
$10^1$-$10^4$ GeV range.
Other proposed cadidates to DM particles are the  ultralight fuzzy dark matter ($\sim 10^{-31}$
GeV), where wave-like behavior suppresses small-scale structures, sub-GeV sectors including QCD axions and sterile neutrinos and
the trans-Planckian regime of WIMPzillas ranging from $10^{10}$-$10^{15}$ GeV or primordial black holes (PBHs) with masses larger than
$10^{45}$ GeVr, for a review see \cite{bertone2018,bozorgnia2024}.
It has been already predicted that compact astrophysical objects can capture
and retain dark matter  from their environments, potentially leading to 
hybrid configurations in which baryonic and DM coexist. Early studies 
already pointed out that the deep gravitational potential wells of neutron stars and 
other compact remnants make them candidates for dark matter accumulation 
through scattering and capture processes, even in scenarios where the interaction 
between the two sectors is weak or purely gravitational \cite{goldman1989}. 
Subsequent works have explored this possibility, demonstrating that 
captured dark matter can modify the internal structure, stability, 
and macroscopic observables of compact objects, including their maximum mass, radius, 
and cooling properties \cite{kouvaris2008,bertone2008}. More recent developments 
employing two-fluid descriptions have shown that a dark-matter admixture can give rise 
to stable equilibrium configurations that would otherwise be forbidden within purely 
baryonic equations of state, and may even populate observationally regions 
such as the neutron-star--black-hole mass gap \cite{nelson2019,vikiaris2025}. 

Even in the low-mass regime, recent studies indicate that DM annihilation
can reshape the evolution of objects below the 
substellar boundary. In particular, the associated energy injection can raise the 
minimum mass for sustained hydrogen burning and can produce long-lived, quasi-stationary 
configurations supported by DM annihilation in regions of sufficiently high ambient 
DM density; in this context, the term {\sl dark dwarfs} 
has been introduced for such DM-powered objects \cite{croon2025}. 
Exoplanets and substellar objects also serve as targets for DM searches 
via capture-induced 
heating signatures that correlate with the Galactic DM density \cite{leane2021}. 
Self-consistent treatments of capture and transport further indicate that sufficiently 
interacting DM can form extended, and in some cases surface-peaked, distributions, 
enabling distinct phenomenology across a wide range of planetary and substellar targets 
\cite{leane2023}. 
At the population level, Bayesian hierarchical analyses have been proposed to exploit 
DM-induced heating of exoplanets and brown dwarfs to constrain the Galactic DM density 
profile, with particular sensitivity toward the inner Milky Way \cite{benito2024}. 
Within these heating-based approaches, Acevedo et al.\  showed that dark 
kinetic heating can be sizeable even for low-escape-velocity bodies when long-range 
dark forces generate an effective  dark escape velocity \cite{acevedo2025}. This would enhance kinetic energy 
deposition upon capture and allow present infrared data (e.g., WISE/JWST observations 
of ultra-cold super-Jupiters) to constrain regions of dark-sector parameter space. 
Another strategy targets neutrinos from captured DM in brown dwarfs,
Bhattacharjee and Calore propose using the nearby brown-dwarf population as a collective 
(“stacked”) neutrino target for IceCube/IceCube-Gen2, specifically in long-lived-mediator 
dark-matter scenarios \cite{bhattacharjee2024}.
Conversely, Ilie et al.\ have recently stressed that DM evaporation can 
suppress long-term accumulation in substellar objects,  especially in the sub-GeV regime \cite{ilie2024}.
Thus, capture can enter a 
geometric-saturation regime leading to a DM dependent critical temperature above which 
these objects lose sensitivity as DM probes.
On the other hand,  a number of studies show that efficient accumulation of dark matter is possible
well beyond the assumptions traditionally 
adopted in capture calculations, 
motivating the study of DM components
inside brown dwarfs. Leane and Smirnov 
modeled dark-matter capture accounting for multiscatter regimes,
reflection effects, and diffusion 
inside astrophysical objects \cite{leane2023b}. A result of their analysis is that the 
saturation cross section does not, in general, correspond to maximal capture, 
especially in objects with moderate escape velocities such as brown dwarfs. Instead, 
they showed that interacting dark matter can approach the geometric capture 
limit through repeated scatterings, even when single-scatter capture is inefficient. 
Owing to their relatively large escape velocities compared to planets and their long 
lifetimes, brown dwarfs were shown to be effective at retaining a significant 
fraction of the incident dark-matter flux, supporting the plausibility of large accumulated 
dark-matter populations over gigayear timescales.
Dasgupta et al.\ extended the dark-matter capture formalism to encompass interactions 
mediated by arbitrarily light mediators, explicitly incorporating the effects of 
dark-matter self-interactions \cite{dasgupta2020}. Their work demonstrated that the standard assumption 
of a uniform energy-loss distribution, equivalent to the contact-interaction limit, breaks 
down for light mediators, weakening existing astrophysical constraints 
on DM--baryon scattering. However, this weakening of constraints 
does not inhibit dark-matter accumulation. Instead, they showed that repulsive 
self-interactions have only a minor impact on the total capture rate while 
modifying the internal equilibrium and gravitational collapse criteria of the captured 
dark component. Then, large and stable dark-matter populations can persist 
inside stellar and substellar objects without triggering collapse, even when baryonic 
interactions are weak.

Our approach focuses on the gravitational and structural impact of a non-annihilating 
dark matter component embedded within the object, allowing us to isolate hydrostatic 
and equilibrium effects independently of any additional energy injection mechanisms. 
We focus on self-interacting dark matter (SIDM), which has been proposed 
as an extension of the standard cold dark-matter paradigm \cite{spergel2000}, motivated 
by small-scale structure issues in galaxies since it can solve some problems as the 
core-cusp in dwarf galaxies, the missing satellites in the Local Group 
\cite{tulin2013,tulin2018,sanchezalmeida2025} and the final parsec problem 
\cite{tiruvaskar2025}, among others. If SIDM can be captured and retained inside 
astrophysical objects, it may alter their internal structure. In scenarios where dark 
matter is efficiently captured and self-interacting, the dark component may form a compact 
core whose central density can become comparable to that of the baryonic fluid. 

This paper is organized as follows.
In section 2, we establish a self-consistent two-fluid framework to model the internal structure of self-gravitating brown dwarfs,
treating them as a composite system of a baryonic fluid and a SIDM fluid. This involves solving the
coupled Lane-Emden equations under hydrostatic equilibrium, incorporating distinct equations of state for both components, to
determine how the presence of a dark matter core modifies the baryonic density profiles and mass-radius relations.
In section 3, we extend this framework to analyze the dynamical propertie variations by
calculating the tidal Love numbers and how affect to the secular apsidal precession rate.

\section{Two-Fluid Model}
\label{sec:model_justification}
In our model
we consider brown dwarfs composed of two gravitating fluids: a baryonic component and 
a SIDM component, each treated as a continuous, self-gravitating medium with 
its own equation of state (EoS). We consider that the two fluids interact only through gravity. 
In hydrostatic equilibrium, each component satisfies its own force-balance 
equation, sourced by the total gravitational potential generated by the combined mass 
distribution. This leads to a coupled system of Lane-Emden equations in which the density 
and pressure profiles of the two fluids are mutually constrained.

\subsection{Baryonic equation of state}

In the mass range of 
interest for brown dwarfs, the interior is close to adiabatic over most of its extent 
and supported by a combination of thermal pressure and partial electron degeneracy; both 
effects are well captured by a polytropic relation with a 
suitably chosen index. Then, we adopt the following equation of state for the 
baryonic fluid,
\begin{equation}
P_b = K_b \rho_b^{1+1/n},
\end{equation}
with polytropic index $n=3/2$ and
\begin{equation}
K_b = \frac{(3\pi^2)^{2/3}}{5}\,
\frac{\hbar^2}{m_e}
\frac{1}{(\mu_e m_u)^{5/3}}.
\end{equation}
For a hydrogen--helium mixture with approximately solar composition, appropriate for 
brown dwarfs, one has $\mu_e \simeq 1.15$ \cite{auddy2016}, yielding
\begin{equation}
K_b \simeq 8.9 \times 10^{12},
\end{equation}
in cgs units. This value is consistent with standard treatments of non-relativistic 
degenerate interiors and provides a good approach to the mass--radius relation for isolated 
brown dwarfs,
which remains valid for brown-dwarfs with masses above $\sim 19\,M_{\rm J}$, 
where Coulomb corrections remain subdominant \cite{croon2025}. Moreover, in the 
presence of an additional non-standard energy source this Coulomb-driven breakdown 
is shifted to even lower masses, so the $n=3/2$ polytropic description is safely 
applicable across the mass range we are going to consider:  $\sim 30$--$80\,M_{\rm J}$. 
We therefore regard the polytropic description of the baryonic sector as a baseline 
model since it is sufficiently flexible to represent the dominant pressure support in the 
relevant mass and density regime.
This approach intentionally bypasses the atmospheric and transport physics detailed in full evolutionary models.
By absorbing these boundary and thermal corrections into an effective constant, $K_b$, our baseline model isolates
the structural and hydrostatic response of the baryonic fluid to the additional gravitational potential of the SIDM core. 
Therefore, our results for structural signatures and tidal deformability represent relative deviations from a baseline, not absolute
observational predictions. The integration of a full
microphysical equation of state is deferred to future precision studies.

\subsection{SIDM equation of state}
In order to model the SIDM component captured inside brown dwarfs, we adopt an 
isothermal EoS:
\begin{equation}
P_{\rm DM} = \sigma_0^2\,\rho_{\rm DM},
\end{equation}
where $\rho_{\rm DM}$ is the SIDM mass density and $\sigma_0$ is the one-dimensional velocity dispersion. The isothermal EoS provides
a useful approximation since SIDM self-interactions drive the phase-space distribution toward a Maxwellian form. This occurs via kinetic
equilibration and establishes a constant velocity dispersion throughout the SIDM core. This effect is independent of the baryonic
thermal state. Furthermore, the isothermal model introduces only one parameter, $\sigma_0$ which links to the velocity
scale of SIDM models.

If baryon–SIDM interactions are negligible and self-interactions dominate, the velocity 
dispersion $\sigma_0$ is set by the virial equilibrium of the SIDM component within the 
total gravitational potential of the system. 
For a self-gravitating system of mass $M$ and characteristic radius $R$, the global virial velocity 
scale is $\sigma_{\rm vir}\sim (GM/R)^{1/2}$. For brown dwarfs with $M\sim 30$--$80\,M_{\rm J}$ 
and $R\sim R_{\rm J}$ lies in the range $\sigma_{\rm vir}\sim 230$--$380~{\rm km\,s^{-1}}$. However, 
for the dark matter to form a compact core embedded deep within the baryonic envelope, 
its effective velocity dispersion must be sub-virial relative to the global potential ($\sigma_0 < \sigma_{\rm vir}$). 
This motivates our choice of  $\sigma_0=50$--$100~{\rm km\,s^{-1}}$. Such values yield 
a centrally concentrated SIDM distribution with a characteristic spatial extent $r_{\rm DM}$ that 
is a fraction of the total radius, allowing us to isolate the structural impact of a dark core 
on the surrounding baryonic fluid.

The adoption of an independent isothermal equation of state for the dark matter component  relies on the assumption of thermal
decoupling between the dark and baryonic sectors. Within the SIDM paradigm, the dark matter
self-interaction cross-section is typically orders of magnitude larger than the scattering cross-section between dark matter
particles and Standard Model baryons. Thus, once captured, the dark matter fluid undergoes rapid kinetic equilibration
primarily through self-scattering. 
Assuming these baryon-dark matter interactions remain subdominant, the timescale for conductive heat exchange between the two
distinct fluids exceeds the internal dynamical relaxation time of the dark core. This physical scale separation ensures that
the dark matter component does not thermalize with the surrounding plasma over gigayear timescales.

\subsection{Hydrostatic equilibrium}
The coupled hydrostatic equilibrium equations read
\begin{align}
\frac{d\Pb}{dr} &= -\rhob \frac{d\Phi}{dr}, \\
\frac{d\PDM}{dr} &= -\rhoDM \frac{d\Phi}{dr},
\end{align}
where $\Pb$, $\PDM$ are the pressures and $\rhob$ and $\rhoDM$ are the densities of the barions and SIDM, respectively.  
The gravitational potential $\Phi$ satisfies
\begin{equation}
\nabla^2 \Phi = 4\pi G (\rhob + \rhoDM).
\end{equation}

We express the baryonic and SIDM densities as functions of the dimensionless variables 
$\theta_b$ and $\theta_{\rm DM}$, respectively,
\begin{equation}
\rho_b=\rho_{b,c}\,\theta_b^n(\xi)\,\,\, {\rm and }\,\,\,\rho_{\rm DM}=\rho_{{\rm DM},c}\,\exp(-\theta_{\rm DM}(\xi)),
\end{equation}
where $r=a\xi$ with $\xi$ an adimensional length, 
$\rho_{\rm DM,c}$ and $\rho_{b,c}$ are the SIDM and baryonic central densities,  respectively, $n$ is the polytropic index 
and $a$ is defined below. Thus, we 
obtain a pair of coupled Lane-Emden equations 

\begin{equation}
\frac{1}{\xi^{2}}\frac{d}{d\xi}\left(\xi^{2}\frac{d\theta_{b}}{d\xi}\right)
= -\left(\theta_{b}^n + \chi\,\exp (-\theta_{\rm DM})\right),
\label{eq:lane-1}
\end{equation}

\begin{equation}
\frac{1}{\xi^{2}}\frac{d}{d\xi}\left(\xi^{2}\frac{d\theta_{\rm DM}}{d\xi}\right)
= \Delta\left(\theta_{b}^n/\chi + \exp(-\theta_{\rm DM})\right),
\label{eq:lane-2}
\end{equation}
defining $\chi$ as the central density ratio
\begin{equation}
\chi \equiv \frac{\rho_{\rm DM,c}}{\rho_{b,c}}.
\label{eq:rho_ratio}
\end{equation}
We have the following boundary conditions at the centre,
\begin{equation}
\theta_b(0)=1,\,\,\,\, \left.\frac{d \theta_b}{d \xi}\right|_{\xi=0}=0,
\end{equation}
\begin{equation}
\theta_{\rm DM}(0)=0,\,\,\,\, \left.\frac{d \theta_{\rm DM}}{d\xi}\right|_{\xi=0}=0.
\end{equation}
The characteristic length scale $a$ appearing in the radial rescaling is given by
\begin{equation}
a^{2} = \frac{(n+1)\,K_b\,\rho_{b,c}^{\,1/n - 1}}{4\pi G},
\end{equation}
and the parameter $\Delta$ in Eq.~\eqref{eq:lane-2} by
\begin{equation}
\Delta \equiv
\frac{(n+1)\,K_b\,\rho_{b,c}^{\,1/n-1}\,\rho_{{\rm DM},c}}
     {\sigma_0^2},
\end{equation}
which measures the relative importance of the SIDM component, combining 
the baryonic equation of state, the SIDM central density, and its velocity dispersion.

\subsection{Feasibility of Dark Matter Core Formation
}
\label{subsec:chi_plausibility}

We now consider whether values as large as $\chi \sim 1$ can be realised within 
gigayear timescales in astrophysical environments. Here we provide an order-of-magnitude 
argument, based on capture and subsequent redistribution of the captured SIDM population, 
that motivates exploring $\chi \sim \mathcal{O}(1)$ as a physically admissible regime in 
sufficiently dense dark-matter environments.
While our analysis treats the central density ratio, $\chi$, as a free parameter, exploring values up to $\chi \sim \mathcal{O}(1)$ 
is physically motivated by
recent advancements in dark matter capture formalism. As demonstrated by \cite{leane2023}, substellar objects can approach
the geometric capture limit through multiscatter processes, particularly for strongly interacting dark matter candidates. 
However, dark matter evaporation can deplete accumulated populations in the sub-GeV mass regime
\cite{ilie2024}
Thus, the $\chi\sim 1$ configurations modeled in this work implicitly assume a macroscopic dark matter population that
either consists of particles sufficiently massive to evade evaporation, or resides in dense Galactic environments where
the capture rate dominates thermal losses over gigayear timescales. Under these favorable regions of the dark sector
parameter space, the geometric concentration of the dark core allows for relevant central densities even if the global
dark matter mass fraction, $f_{\rm DM}=M_{\rm DM}/M_b$, remains subdominant.
It is important to distinguish the global dark-matter mass fraction, 
$f_{\rm DM}$, from the central ratio $\chi$. Even a small 
$f_{\rm DM}$ can correspond to $\chi\sim 1$ if the SIDM distribution is concentrated 
within a radius $r_{\rm DM}\ll R$. A simple scaling can be obtained by relating 
$\rho_{\rm DM,c}$ to the enclosed SIDM mass. If most of the SIDM mass lies within a 
characteristic scale $r_{\rm DM}$, one may write
\begin{equation}
\rho_{\rm DM,c}\sim \frac{3 M_{\rm DM}}{4\pi r_{\rm DM}^3},
\end{equation}
which yields
\begin{equation}
\chi \sim \frac{3}{4\pi}\,
\frac{M_{\rm DM}}{\rho_{b,c} r_{\rm DM}^3}
= \frac{3}{4\pi}\, f_{\rm DM}\,
\frac{M_b}{\rho_{b,c} r_{\rm DM}^3}.
\label{eq:chi_scaling_basic}
\end{equation}
Using $M_b \simeq (4\pi/3)\,\bar\rho_b R^3$, this can also be written as
\begin{equation}
\chi \sim f_{\rm DM}\left(\frac{\bar\rho_b}{\rho_{b,c}}\right)\left(\frac{R}{r_{\rm DM}}\right)^3,
\label{eq:chi_scaling}
\end{equation}
which makes explicit that large $\chi$ can arise from geometric concentration even 
when $f_{\rm DM}\ll 1$.

In our isothermal SIDM approach, the scale $r_{\rm DM}$ is set by hydrostatic balance 
of the SIDM component inside the total potential. A useful estimate is obtained by 
assuming that, in the inner region, the baryons dominate the gravitational acceleration 
so that $g(r)\simeq (4\pi/3)G\rho_{b,c}\,r$. Solving the isothermal equilibrium equation 
then gives an approximately Gaussian SIDM core with scale
\begin{equation}
r_{\rm DM}\sim \left(\frac{3}{2\pi}\right)^{1/2}\frac{\sigma_0}{\sqrt{G\rho_{b,c}}}\,,
\label{eq:rDM_est}
\end{equation}
up to factors of order unity that depend on the detailed two-fluid solution. 
Equation~(\ref{eq:rDM_est}) captures the key trend: smaller $\sigma_0$ (or larger 
$\rho_{b,c}$) implies a more compact SIDM core and, through Eq.~(\ref{eq:chi_scaling}), 
a larger $\chi$ at fixed $f_{\rm DM}$.

\begin{figure}
  \centering
  \includegraphics[width=\columnwidth]{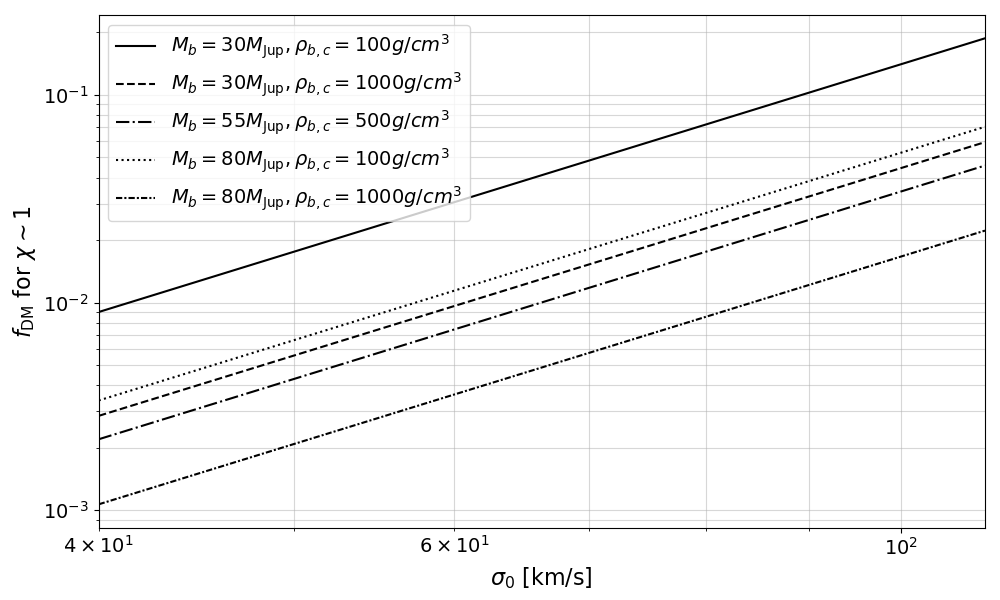}
  \caption{
   Estimated dark matter-baryonic mass ratio, $f_{\rm DM}$ as a function of velocity dispersion $\sigma_0$ for a number of values of total baryonic mass and
baryonic central density for $\chi\sim 1$.
}
  \label{fig:fdm_versus_sigma0}
\end{figure}

To illustrate magnitudes, take representative central baryonic densities in the 
brown-dwarf regime (typically $\rho_{b,c}\sim 10^2$--$10^3~{\rm g\,cm^{-3}}$ in our models) 
and $\sigma_0=50~{\rm km\,s^{-1}}$. Equation~(\ref{eq:rDM_est}) then gives 
$r_{\rm DM}\sim 0.05$--$0.15~R_{\rm J}$, so $(R/r_{\rm DM})^3\sim 10^2$--$10^3$. 
With $\bar\rho_b/\rho_{b,c}\sim 0.1$--$0.3$, Eq.~(\ref{eq:chi_scaling}) implies that 
$\chi\sim 1$ can correspond to a global fraction as small as $f_{\rm DM}\sim 10^{-2}$--$10^{-1}$, 
depending on $(\rho_{b,c},\sigma_0)$ as can be seen in Fig.~\ref{fig:fdm_versus_sigma0}.

\subsection{Lane-Emden solutions}
\begin{figure}
  \centering
  \includegraphics[width=\columnwidth]{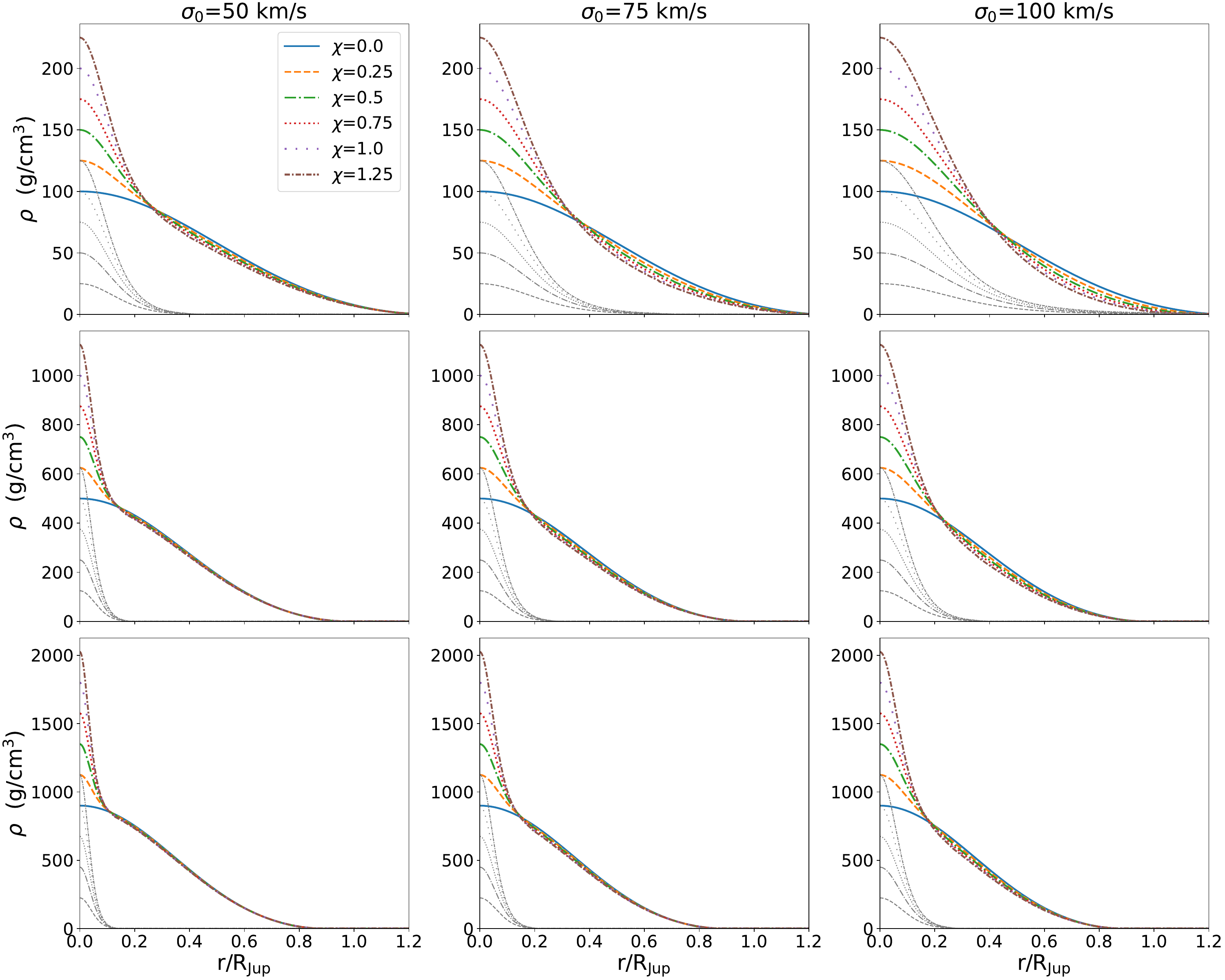}
  \caption{Density profiles for several values of the central density ratio, $\chi$.
  Panels in the same row correspond to the same baryonic central density, while panels
  in the same column share the same SIDM velocity dispersion. Darker lines denote the
  total mass density (baryonic plus SIDM), whereas lighter gray lines correspond to the density of the SIDM component.}
  \label{fig:density_profile.pdf}
\end{figure}
The coupled differential equations \eqref{eq:lane-1} and \eqref{eq:lane-2} are solved 
numerically using a fourth-order Runge-Kutta scheme. Figure~\ref{fig:density_profile.pdf} 
 displays  the density profiles  of our  two-fluid configurations. 
It  shows the total density ($\rhob+\rhoDM$)  and SIDM density profiles for several values of the central density 
ratio, $\chi$, arranged such that panels in the same row share the same baryonic central 
density and panels in the same column share the same SIDM velocity dispersion. Across 
this parameter space the baryonic density sets the baryonic envelope, while the SIDM 
density (lighter gray curves) is centrally concentrated. 
In Figure~\ref{fig:density_profile.pdf} one can check  that 
the total density remains baryon-dominated outside the inner core, and any 
SIDM contribution becomes negligible at radii where the baryonic density still provides 
the bulk of the mass. Motivated by these profiles, we define the brown-dwarf radius 
using the baryonic radius, $R_{\rm b}$, i.e., the radius at which the baryonic density 
vanishes. Since the SIDM component is confined to the central region and 
does not produce an extended outer layer. 

\begin{figure}
  \includegraphics[width=\columnwidth]{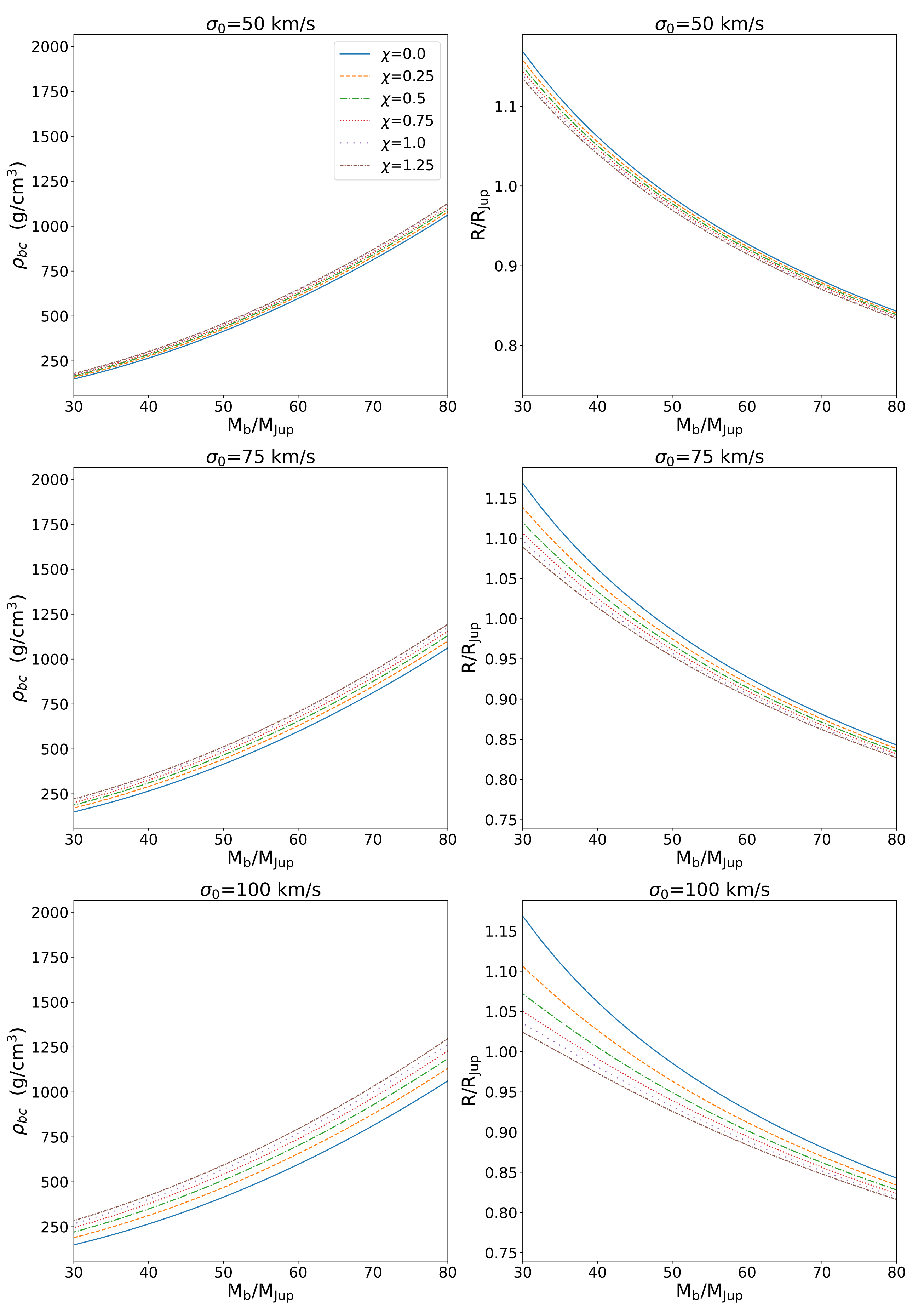}
  \caption{Central baryonic density and radius as functions of the baryonic mass for 
  two-fluid models. The left column shows the central baryonic density as a function 
  of baryonic mass, while the right column displays the corresponding radius-mass 
  relation. Each panel includes several curves corresponding to different values of 
  the central density ratio,  $\chi$, between baryonic matter and dark matter. }
  \label{fig:rho_c_Radius_vs_barionic_Mass}
\end{figure}

\begin{figure}
  \centering
\includegraphics[width=1\columnwidth]{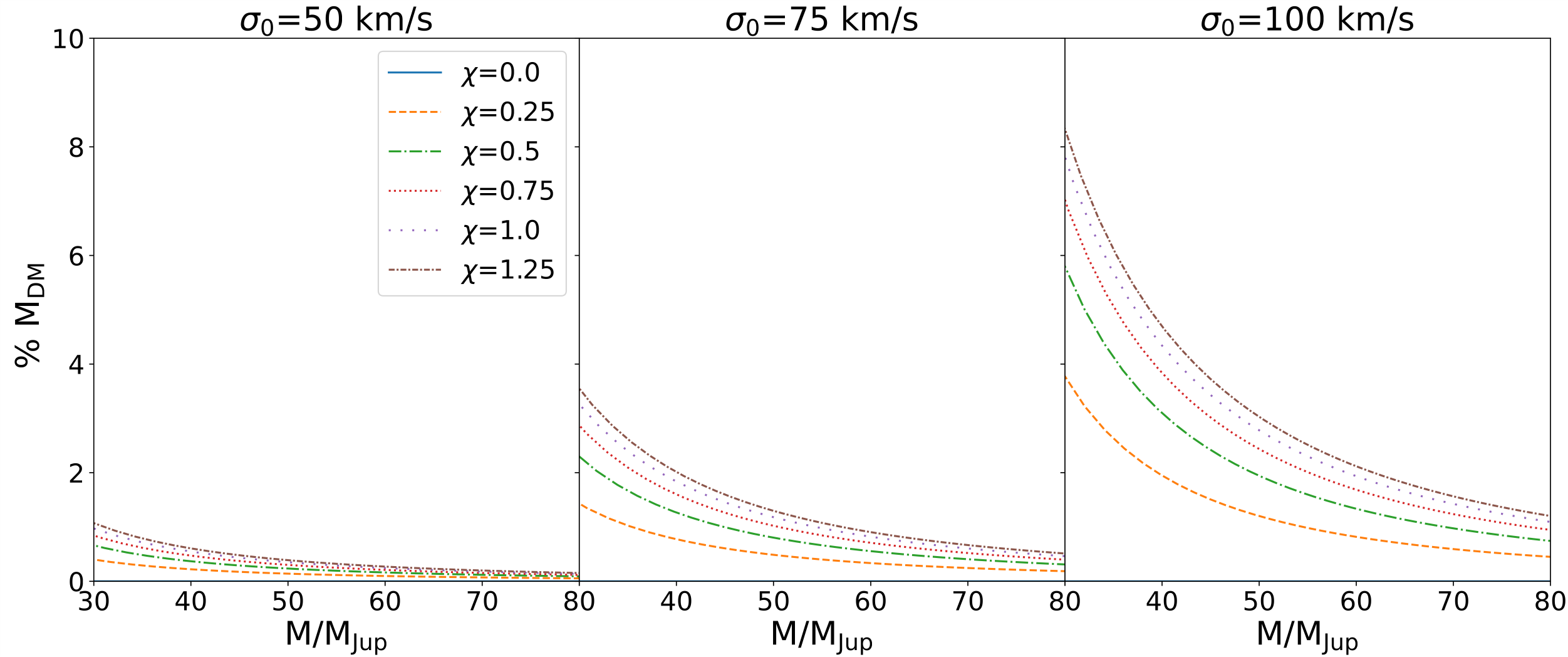}
  \caption{Dark-matter mass fraction as a function of the total mass for two-fluid 
  models. Each panel displays the percentage of dark matter, versus the total mass. 
  Curves correspond to different values of the central density ratio $\chi$ for 
  different SIDM velocity dispersions, $\sigma_0$, as indicated.}
  \label{fig:DM_percent}
\end{figure}

Figure~\ref{fig:rho_c_Radius_vs_barionic_Mass} illustrates how the presence of a
SIDM core modifies the global structure of the baryonic configuration 
along a baryonic-mass sequence. The left column shows the central baryonic density, 
$\rho_{\mathrm{b},c}$, as a function of the baryonic mass, while the right column shows 
the associated radius-mass relation, with rows corresponding to different
values of $\sigma_0$. In each panel, varying the central density ratio $\chi$ 
effectively tunes the relative importance of the dark component: larger $\chi$ enhances 
the gravitational potential in the inner region, requiring the baryonic fluid to adjust 
its central concentration and leading to systematic shifts in both $\rho_{\mathrm{b},c}(M_{\rm b})$ 
and $R(M_{\rm b})$. 
Throughout the parameter space explored, the SIDM component remains a subdominant 
contributor to the overall mass budget. For most choices of $\chi$ and $\sigma_0$, the 
fraction $M_{\rm DM}/M_{\rm tot}$ stays at the level of a few percent or below, 
consistent with the picture of a compact SIDM core embedded within an otherwise 
baryon-dominated brown dwarf, as shown in Fig.~\ref{fig:DM_percent}. Only in the most 
extreme configurations considered, corresponding to the largest values of $\chi$ 
(largest central dark density) and the most favorable $\sigma_0$, does the accumulated 
SIDM content become appreciable, reaching at most $\sim 10\%$ of the total mass. 

\begin{figure}
  \centering
  \includegraphics[width=1\columnwidth]{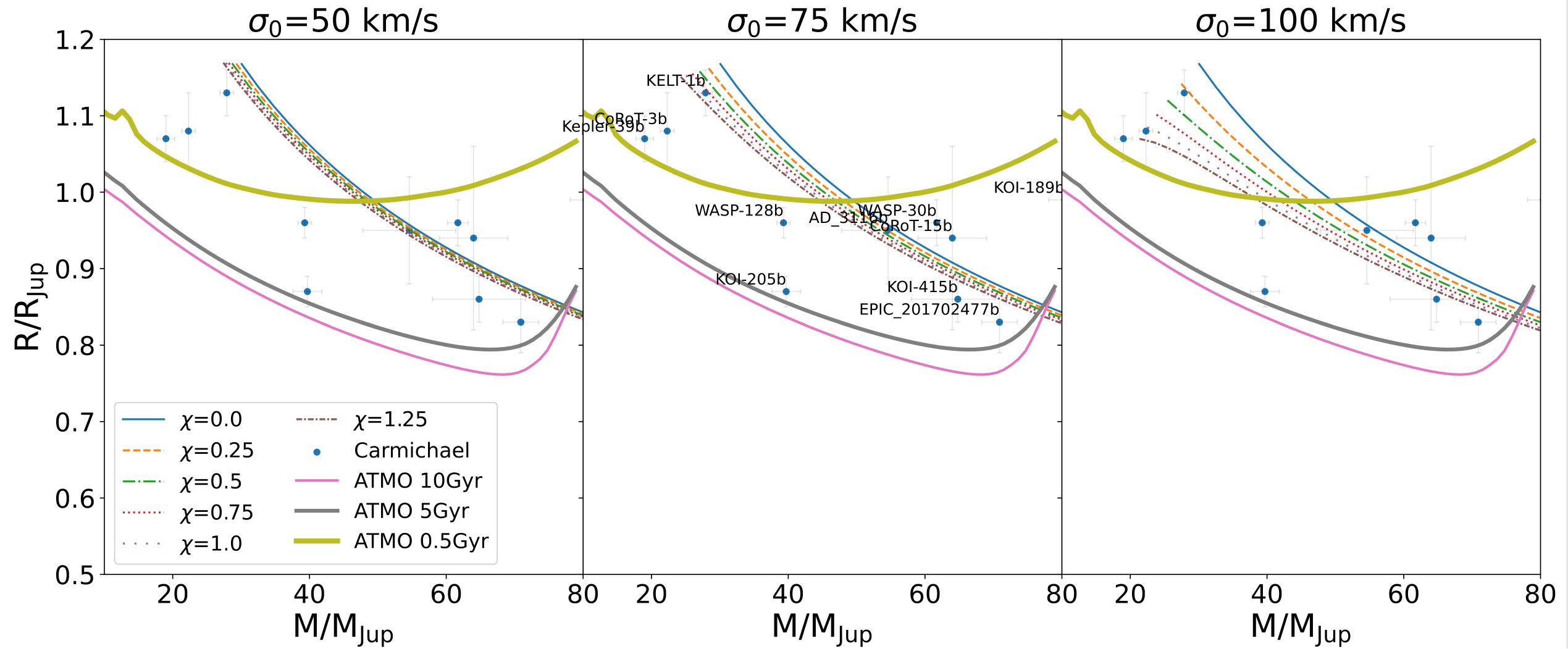}
  \caption{Mass-radius relation for two-fluid brown dwarf models including a SIDM 
  component. Each column corresponds to a different value of the SIDM velocity 
  dispersion, $\sigma_0$, as indicated at the top of the panels. For comparison, 
  ATMO evolutionary models at ages of 0.5, 5, and 10 Gyr are overplotted. Observationally 
  inferred masses and radii from the Carmichael sample are shown as filled symbols 
  \cite{carmichael2023}. The inclusion of a SIDM component leads to systematic 
  deviations from standard evolutionary tracks, whose magnitude depends on $\sigma_0$ 
  and becomes more pronounced at lower masses.}
  \label{fig:Radius_Mass}
\end{figure}

The mass--radius relations obtained from evolutionary models, such as COND 
\cite{baraffe2003,baraffe2015} and ATMO \cite{phillips2020}, differ quantitatively 
from those derived within our polytropic framework. As discussed before,
this discrepancy is expected. State-of-the-art 
evolutionary models incorporate temperature-dependent effects, entropy variations, and 
complex atmosphere-interior coupling to reproduce detailed observables like effective 
temperatures and spectra. In contrast, our polytropic approach intentionally absorbs 
these corrections into an effective constant to isolate the 
leading-order hydrostatic balance of the degenerate interior. 
Figure~\ref{fig:Radius_Mass} shows the mass--radius relation obtained from our two-fluid 
models for several values of the SIDM velocity dispersion, with each panel corresponding 
to a fixed $\sigma_0$, and for different central dark-to-baryonic density ratios. 
The eleven transiting brown dwarfs with the most precise radius determinations are 
overplotted \cite{carmichael2023}, together with ATMO evolutionary tracks at different ages. 
As discussed by Carmichale \cite{carmichael2023}, even for transiting systems the uncertainty on the brown-dwarf 
radius is often dominated by the stellar radius determination, and the incorporation 
of Gaia~DR3 parallaxes \cite{vallerani2023} has led to significant revisions of 
several benchmark objects, in some cases at the $>2\sigma$ level. In particular, 
a number of brown dwarfs previously thought to populate the very small-radius, old 
regime have moved to systematically larger radii, leaving this region of the mass--radius 
plane comparatively sparse. When compared with standard ATMO predictions, several of 
the measured radii therefore depart from the expectations of simple single-component 
evolutionary models at fixed age, highlighting both residual observational systematics 
and potential missing physics in the interior modelling. Our 
SIDM-augmented models naturally introduce an additional degree of freedom that broadens 
the theoretical mass--radius locus and can accommodate part of the observed scatter, 
while remaining consistent with the revised radii reported by Carmichael \cite{carmichael2023}.

\section{Tidal deformability and Love numbers in two-fluid configurations}
\label{sec:love}

The response of a self-gravitating object to an external tidal potential is commonly 
characterised by the Love numbers, which quantify the induced multipolar deformation 
of the gravitational potential. The second-order Love number $k_2$ 
provides a direct measure of the internal mass distribution and central concentration 
of the object, shedding light of its internal structure. Originally 
introduced in the context of terrestrial tides \cite{love1911}, Love numbers have 
since been widely applied to stars and planets, where they enter naturally in the 
theoretical description of tidal and rotational contributions to apsidal motion 
\cite{brooker1955}. 
To compute the tidal Love numbers, we introduce a static, external tidal field and 
study linear perturbations of the gravitational potential. In the Newtonian regime, 
the relevant perturbation equation for the $\ell$ mode can be written as \cite{yip2017}:
\begin{equation}
\frac{d^2 H}{dr^2} + \frac{2}{r}\frac{dH}{dr}
- \left[ \frac{\ell(\ell+1)}{r^2} - \frac{4\pi G\rho^\prime(r)}{\Phi^\prime(r)} \right] H = 0 ,
\label{eq:love_ode}
\end{equation}
where $H(r)$ encodes the radial dependence of the potential perturbation and primes 
denote derivatives with respect to $r$. Regularity at the origin requires $H(r) \propto r^2$ 
as $r \to 0$, while the solution is matched at the surface of the object to the external 
tidal potential. 

Instead of solving \eqref{eq:love_ode}, we solve the differential equation for
\begin{equation}
y(r) \equiv  \frac{r}{H}\frac{dH}{dr},
\end{equation}
that can be easily computed from \eqref{eq:love_ode} as:
\begin{equation}
y^\prime=\frac{\ell(\ell+1)-y^2-y}{r}-\frac{4\pi r^3 \rho^\prime}{m(r)},
\label{eq:love2_ode}
\end{equation}
where $m(r)$ is the mass enclosed within radius $r$. We switch from the differential 
equation \eqref{eq:love_ode} to the equivalent first-order equation \eqref{eq:love2_ode} 
because it is numerically more robust (avoiding cancellations and stiffness in $H$ 
and its derivatives near the center and the surface) while yielding the same Love 
number through a simpler and more stable surface matching condition. It is also well 
known that \eqref{eq:love2_ode} could contain divergent terms for polytropic indices 
between 0 and 1 \cite{yip2017}, a range that does not apply to our study.

The Love number $k_\ell$ is given by
\begin{equation}
k_\ell = \frac{\ell - y(R)}{\left[ \ell+1 + y(R) \right]}.
\end{equation}
Here $R$ denotes the radius of the baryonic component, which defines the effective 
surface of the object for tidal interactions. The SIDM component may extend beyond 
or remain confined within this radius, depending on its velocity dispersion and 
central density, but its gravitational contribution is fully accounted for through 
$\rho(r)$.

\begin{figure}
  \centering
  \includegraphics[width=1\columnwidth]{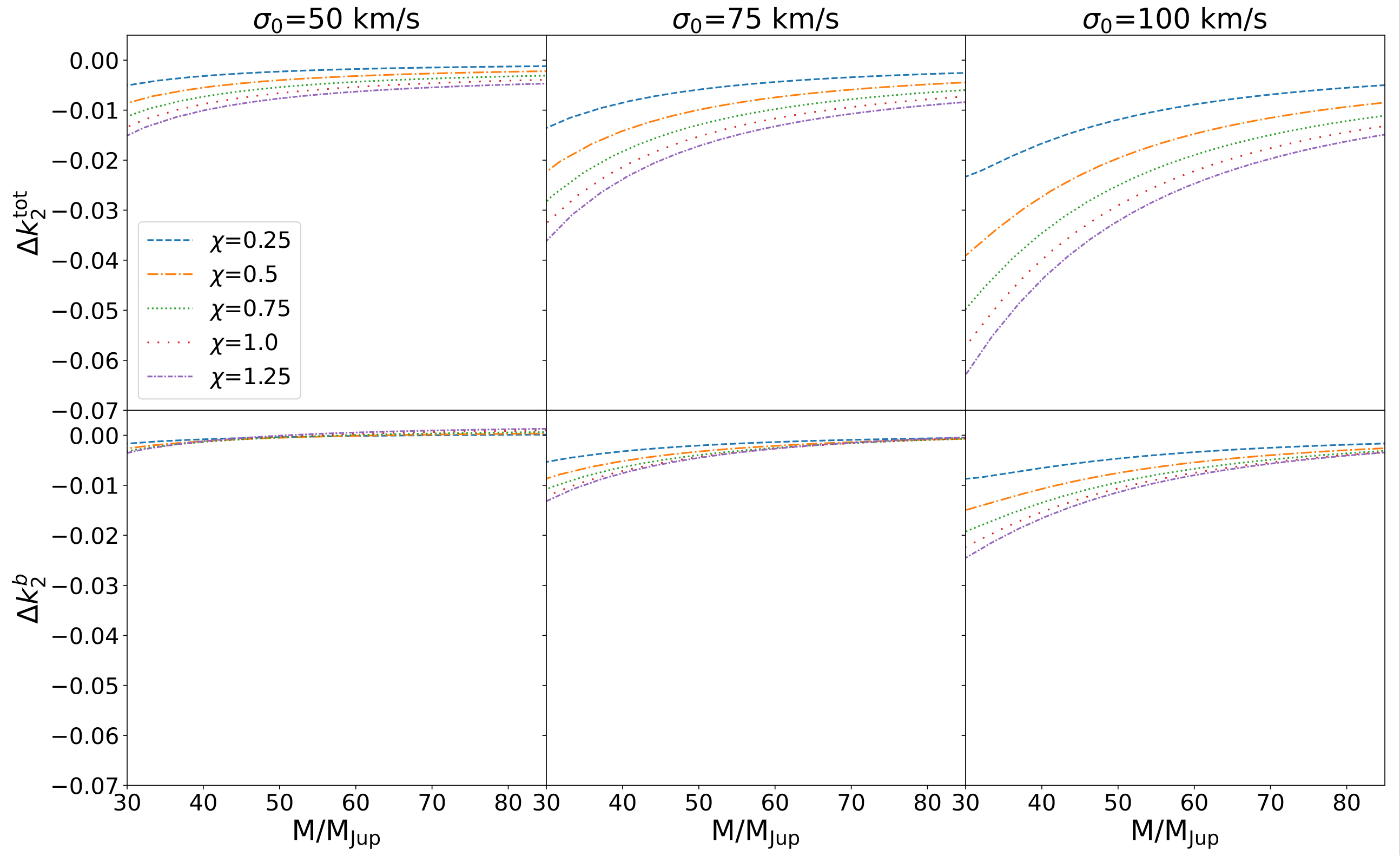}
  \caption{Predicted variations in the Love number $\Delta k_2$ 
  as a function of the brown dwarf mass. The top panels show the variation calculated using  the total density ($\rhob+\rhoDM$),
while the bottom
  panels show the variation calculated using only the baryonic density ($\rhob$), which is affected by the presence of the SIDM component.
} 
  \label{fig:Delta_Love}
\end{figure}

In two-fluid configurations, the Love number depends not only on the total mass and 
radius, but also on the relative central density ratio $\chi = \rho_{\rm DM,c}/\rho_{b,c}$ 
and on the SIDM velocity dispersion $\sigma_0$, which controls the spatial extent 
and compressibility of the dark component. 
Figure~\ref{fig:Delta_Love} shows the variation in the Love number due to the presence of a SIDM core. The top and bottom panels present $\Delta k^{\rm
tot}_2=k^{\rm tot}_2-k_2^0$ and $\Delta k^{\rm b}_2=k^{\rm b}_2-k_2^0$, respectively. $k_2^0$ is the Love number in the
absence of SIDM for a given mass, $k^{\rm tot}_2$ and $k^{\rm b}_2$  are the Love number derived  from the total density
(baryonic$+$SIDM) and the baryonic only density profile, respectively.
For fixed total mass, increasing the SIDM 
fraction generally leads to more centrally concentrated configurations, resulting in 
smaller values of $k_2$ compared to purely baryonic models.
This sensitivity allows tidal observables to break degeneracies present in mass--radius 
relations and provides an additional test to constrain the presence and properties 
of SIDM in substellar objects:  two objects with identical masses and 
radii but different internal SIDM distributions can exhibit measurably different 
tidal responses.

\subsection{Observational implications}

In sub-stellar objects tidal interactions are relevant in close binary 
systems, where they contribute to apsidal 
precession. In transiting systems, 
measurements or constraints on apsidal precession can therefore be translated into 
bounds on $k_2$, offering a measure of the interior structure that is complementary to 
mass--radius determinations \cite{ragozzine2009,kramm2011,kramm2012,bernabo2024}. 
Unlike global quantities such as the total mass or radius, the Love numbers are 
sensitive to the radial density profile and therefore offer a diagnostic of 
additional gravitating components beyond ordinary baryonic matter.
The apsidal precession can be decomposed into the sum 
of three contributions:
\begin{equation}
\frac{d\omega}{dt} \equiv \dot{\omega}
= \dot\omega_{\rm GR}
+ \dot\omega_{\rm tidal}
+ \dot\omega_{\rm rot} .
\end{equation}
Where $\dot\omega_{\rm GR}$ is the general relativity contribution \cite{einstein1915}, $\dot\omega_{\rm tidal}$ and
$\dot\omega_{\rm rot}$  
 the tidal and rotational contribution to the apsidal precession, respectively \cite{sterne1939,csizmadia2019,bernabo2024}.

Using high-precision 
eclipse timings from \textit{TESS}, has been  demonstrated that 
accurate determinations of apsidal motion rates allow stringent tests of stellar 
structure and evolutionary models over a wide range of stellar masses \cite{baroch2021,claret2021}, once the 
relativistic contribution is properly accounted for. Their analysis highlights the 
potential of apsidal precession measurements to complement traditional mass--radius 
constraints and to reveal subtle differences in internal density concentration. The 
impact of stellar rotation on the apsidal motion constants has been quantified in 
detail by \cite{claret2024}, who showed that rotational effects introduce systematic, 
mass-and evolutionary-stage–dependent corrections to $\log k_{2,\rm aps}$ that can 
be consistently included in theoretical apsidal motion calculations. Although direct 
measurements of Love numbers for sub-stellar objects remain challenging, ongoing and 
future high-precision timing and orbital surveys may provide indirect constraints, 
motivating the inclusion of tidal diagnostics in models of dark-matter-admixed objects. 
Note that the Love number, $k_2$ is the double of the apsidal number $k_{2,\rm aps}$ \cite{csizmadia2019}, 
which, for a polytrope of index $n=3/2$, is equal to 0.14327923 \cite{brooker1955}.

In our model, both components contribute to the total gravitational potential and, therefore, to 
the response of the system to tidal perturbation. For this reason, 
the Love number entering the tidal contribution to the apsidal motion rate, 
$\dot\omega_{\rm tidal}$, must be computed from the full mass distribution 
(baryons + SIDM), $k_2^{\rm tot}$. On the other hand, 
the dark matter component could carry no net angular momentum, provided that the 
absence of non-gravitational interactions with the baryonic fluid could not exist an efficient 
mechanism for angular-momentum transfer or spin-up of the dark matter. Any rotational 
motion induced indirectly through the time-averaged gravitational potential is expected 
to be negligible and does not lead to a centrifugal deformation of the dark matter 
distribution. Under this assumption, the Love number controlling $\dot\omega_{\rm rot}$ 
is computed from the baryonic density profile alone, which is also modified by the presence of the SIDM core, 
and we denoted it by $k_2^{b}$.
However,  if  the interaction SIDM-baryons is not negligible, the dark core could posses a net angular moment and the  
Love number would shift from $k_2^b$  toward $k_2^{\rm tot}$.
In Fig.~\ref{fig:Delta_Love} the variation of $k_2^{\rm tot}$ and $k_2^b$ is shown for a number of values of $\chi$ as a function of mass.
Then, the  total apsidal precession can then be written as
\begin{equation}
\dot\omega \;=\; \dot\omega_{\rm GR}
\;+\; \dot\omega_{\rm tidal}\!\left(k_2^{\rm tot}\right)
\;+\; \dot\omega_{\rm rot}\!\left(k_2^{r}\right),
\end{equation}
where $k^r_2$ is between $k_2^b$ and $k_2^{\rm tot}$, being equal to $k_2^b$ when the SIDM-baryon interaction is negligible (assuming the SIDM core is not rotating), or equal to $k_2^{\rm tot}$ if the
SIDM and the baryonic fluid are corotating.
Using the equations for $\dot{\omega}$ from \cite{csizmadia2019}, the variations induced by the presence of SIDM can be written as 
\begin{equation}
\Delta\dot\omega_{\rm tidal}
\simeq
\frac{15}{2}\,
\frac{n_m}{\left(1-e^{2}\right)^{5}}
\frac{M_\star}{m_{\rm BD}}\,
\Delta k^{\rm tot}_{2}
\left(\frac{R_{\rm BD}}{a}\right)^{5}
\left(
1 + \frac{3}{2}e^{2} + \frac{1}{8}e^{4}
\right) .
\label{eq:omega_tide}
\end{equation}

\begin{equation}
\Delta\dot\omega_{\rm rot}
\simeq
\frac{1}{2}\,
\frac{n_m}{\left(1-e^{2}\right)^{2}}
\left(\frac{P_{\rm orb}}{P_{\rm rot,BD}}\right)^{2}
\Delta k^r_2
\left(\frac{R_{\rm BD}}{a}\right)^{5}
\left(1+\frac{M_\star}{m_{\rm BD}}\right) .
\label{eq:omega_rot}
\end{equation}
Here $n_m = 2\pi/P_{\rm orb}$ denotes the Keplerian mean motion of the orbit,
$P_{\rm orb}$ is the orbital period and  $M_\star$ and $e$ are the mass of the star and the orbital eccentricity, respectively.
The parameters $m_{\rm BD}$, $R_{\rm BD}$,
and $P_{\rm rot,BD}$ represent the mass, radius 
and rotational period of the brown dwarf, while $a$ is the semi-major axis of
the binary orbit and $\Delta k^i_{2}=k_2^i-k_2^0$ is the variation of Love number due to the SIDM presence, $k_2^0$ is the 
Love number for a SIDM free brown dwarf of the same mass.

\subsection{Measuring apsidal precession}
Continuous monitoring of transit timing variations (TTVs) and radial velocity curves in highly eccentric, short-period systems enables the 
measurement of apsidal precession in transiting binaries.
NASA’s TESS mission \cite{ricker2015} has significantly increased the census of 
transiting brown dwarfs, growing the count from 16 to more than 50 systems 
\cite{vowel2025}. In \eqref{eq:omega_tide} and \eqref{eq:omega_rot} we can check 
that the ratio of the BD radius to the separation $a$ is a critical parameter as it 
is elevated to the fifth power. In order to be able to detect variations in the 
apsidal precession we need transiting binaries with $R_{\rm BD}/a$ as large as possible.
The detectability of this signal requires $\Delta \dot{\omega}$ to exceed the empirical uncertainty of transit timing variation
(TTV) measurements over a given observational baseline. Based on the precision achieved in long-term monitoring of hot Jupiter
systems (e.g., WASP-19b), we adopt an observational threshold of $\Delta \dot{\omega}_{obs} \approx 30^{\prime\prime} \text{
day}^{-1}$.
By setting $\Delta \dot{\omega} = \Delta \dot{\omega}_{obs}$, we can solve for the critical orbital separation $a_{crit}$ that
defines the detectability wall. Substituting the Keplerian mean motion into \eqref{eq:omega_tide} and \eqref{eq:omega_rot}, the total functional dependence of the
signal on the semi-major axis is isolated as:
\begin{equation}
\Delta \dot{\omega} \propto a^{-3/2} \cdot a^{-5} = a^{-6.5}.
\end{equation}
This extreme power-law dependence governs the topology of the detectability wall in the $(a/R_{BD}, q)$ parameter space. Rearranging
for the critical separation yields:
\begin{equation}
\frac{a_{crit}}{R_{BD}} \propto \left( \frac{\Delta k_{2,BD}}{\Delta \dot{\omega}_{obs}} \right)^{1/6.5}.
\label{eq:a_crit}
\end{equation}

\begin{figure}
  \centering
  \includegraphics[width=1\columnwidth]{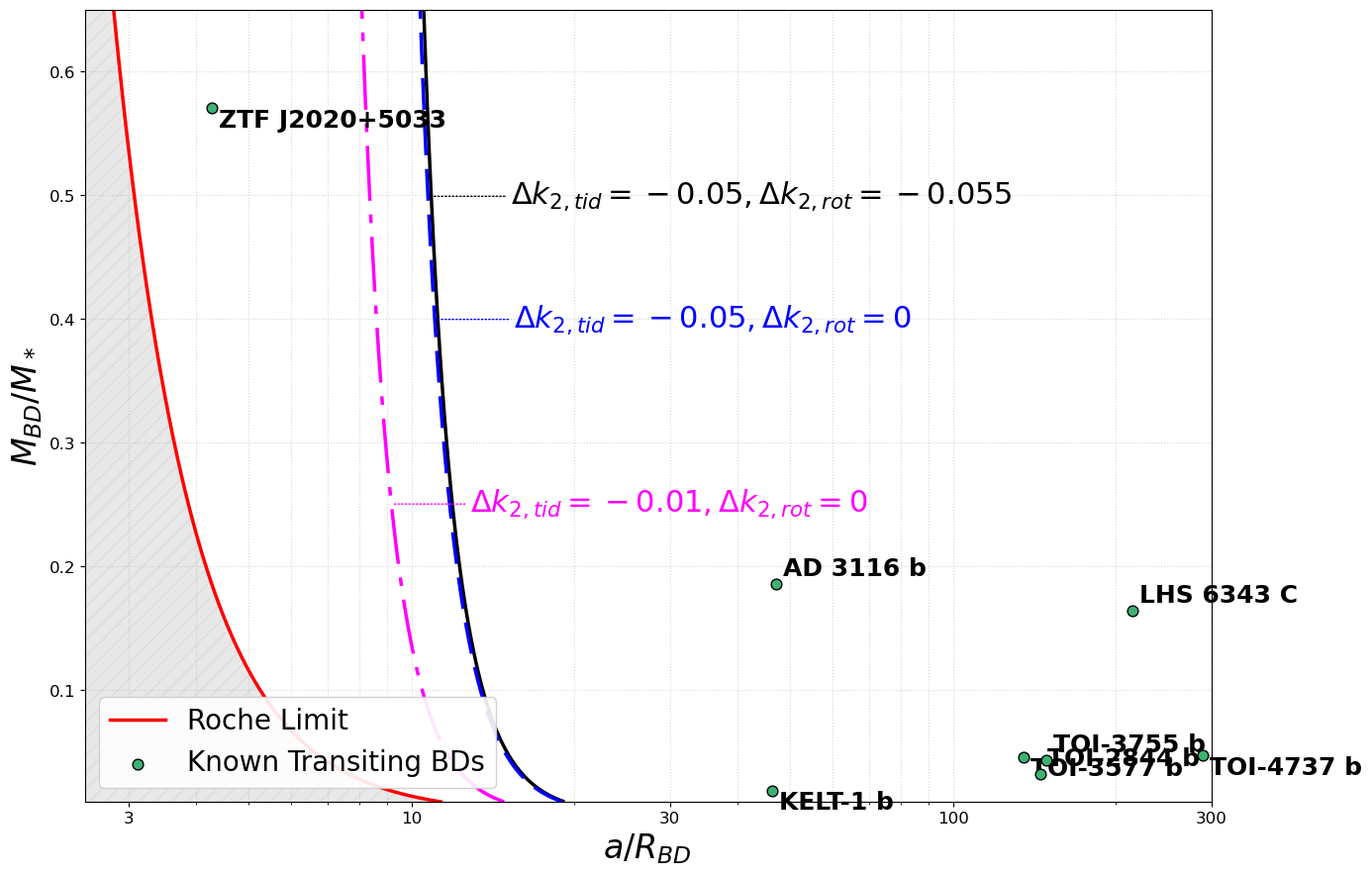}
  \caption{
Detectability map of the periastron precession anomaly ($\Delta \dot{\omega}$) in the parameter space defined by the orbital separation ($a/R_{BD}$) and the mass ratio ($M_{BD}/M_*$).
We assume a fixed eccentricity of $e=0.1$ and that the brown dwarf is in synchronous rotation with the orbital period. The contours denote an empirical observational threshold of $30^{\prime\prime}\text{
day}^{-1}$ for different decoupled scenarios of the brown dwarf's interior
response. }
  \label{fig:Detectability_Map}
\end{figure}
Equation \eqref{eq:a_crit} shows the physical rigidity of the detectability boundary. An attempt to measure a finer variation in the internal
structure, such as decreasing the target $\Delta k_{2,BD}$ from $0.05$ to $0.01$ (a factor of 5), only shifts the maximum
observable orbital distance by a factor of $5^{1/6.5} \approx 1.28$. Consequently, the boundary appears as a nearly vertical
wall across all mass ratios ($M_{BD}/M_*$). In Fig.~\ref{fig:Detectability_Map} we plot these boundaries for several values of $\Delta k_2$ and along with data from some known transiting BDs 
binaries \cite{elbadry2023,carmichael2023,vowel2025}.
 The steepness of this gradient dictates that high-precision constraints on brown
dwarf equations of state via apsidal motion are strictly confined to ultra-short period systems residing exceptionally close to
their Roche fluid disruption limit:
\begin{equation}
\frac{a}{R_{BD}}\approx 2.44\left(\frac{M_*}{M_{BD}}\right)^{1/3}
\end{equation}
This
explains why the vast majority of known transiting brown dwarfs, which typically orbit at dozens of stellar radii, remain unsuitable
for tidal structure studies. In contrast, systems like ZTF J2020+5033,
 a system consisting of a low-mass M dwarf ($0.134 M_\odot$)
and a transiting high-mass brown dwarf ($80.1 M_J$) in an tight 1.90-hour
orbit\cite{elbadry2023},
 which orbit at the very edge of the Roche limit, provide the
only environment where the tidal signal is large enough to distinguish between different internal structure models within reasonable
observational baselines. Although this system in particular is affected by magnetic activity in the primary star \cite{elbadry2023,applegate1992}.

\section{Discussion}
We find that brown dwarfs could be employed as probes to the presence 
of captured SIDM. Unlike 
main-sequence stars, where any effect is 
smeared out by thermal pressure, the degenerate interiors of brown dwarfs are
sensitive to gravitational perturbations induced by a SIDM core.
Our two-fluid Lane-Emden modelling show that these structural deviations appear 
once the dark-matter density in the core becomes relevant 
($\chi \sim 1$). Our analysis shows a strong dependence on
the SIDM velocity dispersion: when the dark matter is in a 
sub-virial thermal state relative to the global potential of the object 
($50 \lesssim \sigma_0 \lesssim 100~{\rm km\,s^{-1}}$), it naturally sinks to 
form a  gravitationally central core. Conversely, for higher 
velocity dispersions ($\sigma_0 > 100~{\rm km\,s^{-1}}$), the SIDM
distribution becomes too hot and diffuse to form such a core, 
thereby diminishing its structural impact. 
Forming this compact dark core directly alters the baryonic hydrostatic equilibrium.
In order to support the 
additional central gravity without a corresponding increase in thermal pressure, 
the baryonic envelope is forced into a more centrally condensed configuration. 
Thus, the theoretical mass--radius relation is systematically shifted, 
leading to a decrease of the radius by several percent 
(up to $\sim 10\%$ in optimal scenarios). This creates specific regions in the 
mass--radius plane that cannot be accommodated by standard single-fluid 
evolutionary models.
However, relying solely on the mass--radius relation presents observational challenges. 
While high-precision radii from transiting systems (e.g., those from the Carmichael 
sample) show some scatter compared to canonical tracks, these deviations might be 
partially masked by systematic uncertainties in host-star parameters, age determinations, 
or atmospheric prescriptions. 
Dynamical observables, specifically tidal deformability, can break this degeneracy.
We have  demonstrated that the SIDM core reduces the 
second-order Love number $k_2$. By dynamically separating the global response to 
external tides ($k_2^{\rm tot}$) from the baryonic response to rotation ($k_2^r$), 
we showed that the apsidal precession rate ($\dot{\omega}$) in close, eccentric binary 
systems is altered. 
Individual systems like ZTF J2020+5033 
are great targets for testing tidal deformability, but separating the dark matter signal from uncertainties in age, metallicity, or
atmospheric models is difficult along with the modulation in orbital period due to magnetic activity of the primary star.
 A statistical, population-level approach can help break this degeneracy.
By comparing the apsidal motion rates of eclipsing brown dwarf binaries situated in
environments with vastly different ambient dark matter densities (e.g., the inner Galaxy versus the local solar neighborhood), one
could identify systematic offsets in the inferred Love numbers. A statistically significant suppression of $k_2$.
in the high-density population, relative to a low-density control sample of similar mass and age, would provide compelling
evidence for SIDM accumulation. This population-wide diagnostic would effectively bypass the need for perfectly calibrated
theoretical baselines for individual substellar objects.

\section{Conclusions}
We have developed a two-fluid model to investigate how the capture of
SIDM redefines the internal physics of brown dwarfs. Our study 
demonstrates that SIDM accumulation shifts the mass--radius relation.
We find that these modifications are driven by the dark matter's thermal state: at 
sub-virial velocities ($\sigma_0 \lesssim 100~\mathrm{km\,s^{-1}}$), the dark core 
forces the baryonic envelope to contract, suppressing the radius by nearly $10\%$. 
This internal compression further manifests as a reduction in the quadrupolar Love number $k_2$, 
which in turn alters the apsidal precession in close binaries. 
Future photometric surveys will expand the census of eclipsing brown dwarfs. In these systems, combined measurements of radii and
apsidal precession will offer a new way to probe the Galactic DM population.

\section*{Acknowledgements}
The author acknowledges 
the project  PID2022-141915NB-C22 funded
by MCIU/\-AEI/\-10.13039/\-501100011033 and FEDER/UE.

\bibliographystyle{JHEP} 
\bibliography{biblio}

\end{document}